# Microcavity-mediated Coupling of Two Distant Semiconductor Qubits

E. Gallardo[1,a)], L.J. Martínez[2], A.K. Nowak[1], H.P. van der Meulen[1], J.M. Calleja[1], C. Tejedor[1], I. Prieto[2], D. Granados[2], A.G. Taboada[2], J.M. García[2] and P.A. Postigo[2]

[1] *Departamento de Física de Materiales, Universidad Autónoma de Madrid, E-28049 Madrid, Spain.*

[2] *Instituto de Microelectrónica de Madrid, Centro Nacional de Microelectrónica, Consejo Superior de Investigaciones Científicas, Isaac Newton 8, PTM Tres Cantos, E-28760 Madrid, Spain.*

[a)] E-mail address: eva.gallardo@uam.es

## Abstract

Long distance (1.4 µm) interaction of two different InAs/GaAs quantum dots in a photonic crystal microcavity is observed. Resonant optical excitation in the p-state of any of the quantum dots, results in an increase of the s-state emission of both quantum dots and the cavity mode. The cavity-mediated coupling can be controlled by varying the excitation intensity. These results represent an experimental step towards the realization of quantum logic operations using distant solid state qubits.

Efficient quantum information applications require qubits with low decoherence rates, fast manipulation times and easy scalability [1]. These requirements are met by qubits based on electron spins or excitons in semiconductor quantum dots (QDs). Coupling of single semiconductor QD excitons to a microcavity confined electromagnetic mode has several advantages depending on the coupling strength. Weak coupling allows enhanced optical efficiency associated to the exciton decay time reduction by the Purcell effect [2]. In the strong coupling regime, the system presents entangled light-matter states that can be used as building blocks for transmission of quantum information [3], qubit readout [4] and production of entangled pairs by compensation of the natural exciton fine structure splitting [5]. Single QD-cavity coupling has been demonstrated in the past years [6-12], showing interesting cavity-quantum electrodynamics effects. The possibility of using two or more qubits coupled by a single optical microcavity is appealing for it can provide techniques for long-distance, fast interactions between qubits [13-15]. New dynamical phenomena are expected in these systems, which are dependent on the relative energy scales of the coupling between qubits and between qubits and the cavity mode (CM). In randomly distributed QDs samples it is statistically difficult to have two or more QDs both spatially and spectrally coupled to a microcavity mode. Some approaches have been proposed to obtain this type of coupled system [16-18], which rely on the deterministic location of the QD in the cavity [11,19]. Coupling of several QDs to a single cavity mode has been reported as the origin of lasing at very low threshold [20].

In this letter we show that exciton states of two semiconductor quantum dots with large lateral separation interact through a microcavity confined optical mode. Individual coupling of the QDs to the CM is demonstrated by changes in photoluminescence (PL) emission intensity and polarization as the QD excitons are brought into resonance with the CM. Cavity-mediated inter-QD interaction is demonstrated by photoluminescence excitation (PLE) measurements, in which resonant excitation at the p-state of any of the QDs increases the s-state emission of the other one. The microcavity-mediated interaction of the two QDs can be controlled by varying the excitation intensity, i.e. the photon number in the cavity mode.

The sample consists of a layer of randomly distributed self assembled InAs QDs grown by solid source molecular beam epitaxy. Atomic force microscope images of the QDs before capping show a ring-like shape [21], which is not relevant for the present results. The QDs are located inside a 158 nm thick GaAs slab grown on top of a 500 nm thick AlGaAs sacrificial layer. The QD surface density is $7.5 \times 10^9$ cm$^{-2}$. The QD height average is 2 nm and the average lateral size is 50 nm. A photonic crystal (PC) triangular lattice of holes of 140 nm diameter with a lattice constant of 230 nm was patterned by e-beam lithography and dry etching. Air suspended membranes were realized by sacrificial etching of the underlying AlGaAs layer. The optical cavity is formed by a missing hole in the PC and a slight inward shift of its nearest neighbours truncated holes (*calzone* cavity) [22], as shown in the inset of Fig. 1(b). A Purcell factor around 450 is obtained for a measured quality factor Q = 3000, an emission wavelength of 920 nm, a refraction index of 3.57 and a mode volume around $8 \times 10^{-3}$ µm$^3$. The lowest energy cavity mode is split into two components with orthogonal linear polarizations separated by ~4 meV. Hereafter they will be designated as V and H modes (V corresponds to the long cavity diagonal). PL and PLE spectra of single QDs were taken with a micro-PL setup with a 2 µm spot size using a Ti-sapphire continuous laser. The in-plane position of the spot could be varied in 14 nm steps. The QD exciton detuning from the cavity mode was varied either by changing the temperature or by controlled Xe thin film deposition [23]. The emission spectra in Fig. 1(a,b) show the exciton emission of two QDs (QD1 and QD2) and the cavity mode (CM) of higher energy (V polarization). One observes that QD1 and CM are fully V-polarized, while QD2 has both V and H components. By scanning the microscope objective in the sample plane to optimize the different PL line intensities one can roughly estimate the location of the QDs. We find that QD1 and QD2 are at 0.5±0.15 and 0.9±0.15 µm respectively, away from the CM maximum almost in opposite directions along H direction (see the inset in Fig. 1(b)). The inter-dot distance is 1.4±0.3 µm. Fig. 1(c) shows the crossing of QD1 and CM energies for increasing temperature. No measurable anticrossing is observed (neither in QD2), indicating weak coupling to the cavity. This is not surprising in view of the long distance from the QDs to the cavity center. However, recent theoretical work [24, 25] has shown that, under incoherent pumping, strong coupling can hold in the absence of a visible anticrossing due to decoherence-induced broadening of the lines. The individual QD coupling to the cavity is shown by temperature tuning of the QD excitons into resonance with the CM. Significant intensity increases are observed at zero detuning, as shown in Fig. 1(d) for both QDs. A second evidence of the QD2 coupling to the CM is shown in Fig. 1(e, f): the polarization diagrams show a rotation of the polarization emission of the QD2 exciton towards the CM polarization direction as its detuning is decreased. The polarization degree $P = \frac{I_{max} - I_{min}}{I_{max} + I_{min}}$ remains unchanged near 0.75, indicating that the influence of CM is to enhance the $I_V/I_H$ polarization ratio of QD2 rather than to provide *independent* V polarized photons (in that case, *P* should decrease for decreasing detuning).

So far we have studied the individual coupling of the QDs to the CM. Now we will show how the cavity mediates an effective interaction between them. The emission intensity of the QDs and CM is presented in Fig. 2 as a function of excitation (vertical scale) and emission (horizontal scale) energies. The PLE maxima of QD1 and QD2 are around 40 meV above their emission energies. These maxima correspond to excited states, which we will simply call p-states. The p-states of both QDs contribute to enhance the CM emission, as shown by the double maximum in the CM trace. The faint feature in the QD2 trace at 1.3798 eV excitation energy (white arrow in Fig. 2) indicates enhanced QD2 PL emission upon p-shell excitation of QD1. This "cross-excitation", schematically drawn in the inset of Fig. 2, proves the effective interaction between the two QDs. This effect is shown in Fig. 3 in more detail, where PLE profiles (vertical cuts in Fig. 2) of QD1 and QD2 are presented. The shoulders appearing in the QD2 PLE profiles for excitation at the p-state of QD1 are systematically observed at other detunings (not shown). Moreover, the clear asymmetry of the QD1 PLE spectra to lower excitation energies is indicative of the reciprocal process, i.e. enhanced QD1 emission upon QD2 p-state excitation. So, each QD becomes brighter upon p-shell excitation *in the other* QD. The amount of "cross-excitation" depends on the external pumping power (which determines the photon number in the cavity mode). Fig. 4 shows PLE spectra of QD2 for H polarization displaying an increasing contribution of the QD1 p-state (the shoulder at 1.3815 eV) for increasing excitation power.

The type of cross-excitation just described is not possible in independent QDs being 1.4 microns apart from each other. A rough estimate of the dipole-dipole interaction for CdSe QDs gives 1 meV for an interdot distance R = 5 nm [17]. Assuming a similar value for the dipole matrix element for our InAs QDs, and considering the $1/R^3$ decrease of the interaction, the direct dipole coupling would be in the range of $10^{-8}$ meV for our dots. Consequently, coupling between the two QDs is only possible through their coherent interaction with the cavity mode. The coupling would proceed in the following way: after resonant excitation at the p-state of one of the QDs, the electron-hole pair relaxes incoherently by phonon emission to the s-state of the same dot and becomes coherently coupled to the second QD by the cavity mode. We can represent this process as $p_1 \rightarrow s_1 \Leftrightarrow CM \Leftrightarrow s_2$, where $\rightarrow$ indicates incoherent decay and $\Leftrightarrow$ coherent coupling. A second channel, in which the excited state of the first QD decays incoherently into a cavity photon, exciting then the second QD ($p_1 \rightarrow CM \Leftrightarrow s_2$), would lead to an incoherent inter-dot interaction. Although this second scenario cannot be ruled out, the following result points to the first one as the most probable: the inset of Fig. 4 shows the contribution of the QD1 p-excitation to the QD2 emission (inter-dot excitation, blue triangles) as a function of power excitation, extracted from a two-Gaussian fit of the PLE spectra. The red crosses stand for the QD2 emission intensity excited at its own p-state (intra-dot excitation). The observed trends indicate that the PL emission of QD2 increases sub-linearly upon increasing the excitation intensity at either of the QD p-states. If the inter-dot QD2 emission were due to direct photon injection from CM, a linear trend would be expected, as the CM emission does not saturate. Consequently, the most probable inter-dot interaction mechanism involves the coherent coupling of the s-states by the cavity mode.

The physics of two qubits with no direct coupling but coupled to the same CM can be writen, by means of a Schrieffer-Wolf transformation, in terms of an effective coupling between the qubits giving an effective Hamiltonian [17]:

$$\hat{H} = \frac{\hbar}{2} J (\hat{s}_1^+ \hat{s}_2^- + \hat{s}_1^- \hat{s}_2^+); \quad J = \left( \frac{g_1^2}{\Delta_1} + \frac{g_2^2}{\Delta_2} \right) \quad (1)$$

where $g_i$ is the coupling strength between qubit $i$ and the cavity mode, $\Delta_i$ is the detuning of qubit $i$ with respect to CM, and $s_{i,}^{\pm}$ are raising and lowering operators for qubit $i$. It describes the photon emission of one qubit into the cavity and subsequent absorption by the other qubit, giving rise to an effective cavity-mediated interqubit coupling. The highest g values reported for InAs QDs are in the order of 0.1 meV [11] for QDs located near the cavity center. In our case $g_i$ must be much smaller due to the relatively large distance to the cavity center, so we can assume safely $g_i \ll \Delta_i$ for $0.5 < \Delta_i < 1.5$ meV. As the observed QD emission energies are separated from the CM one by $\Omega_i = 2\sqrt{g_i^2 + (\Delta_i/2)^2}$ we have $\Omega_i \approx \Delta_i$. A rough estimate of the relative couplings can be made by considering the CM electric field strength at each QD site and the relative emission intensities for equal (large) detuning. As $g \propto \vec{p} \cdot \vec{E}$, where p is the QD dipole moment and E is the CM electric field at the QD site, we have:

$$\frac{g_1}{g_2} = \frac{p_1 \cos\alpha_1 E_1}{p_2 \cos\alpha_2 E_2} = \frac{\sqrt{I_{PL}(1)} \cos\alpha_1 E_1}{\sqrt{I_{PL}(2)} \cos\alpha_2 E_2} \quad (2)$$

From Fig. 1(a,e) we have $\alpha_1 = 0°$ and $\alpha_2 = 80°$. The dipole moments are taken to be proportional to the square root of the QD maximum PL emission intensity $I_{PL}(1,2)$. The PL intensities are measured for equal (and as large as possible) detunings and normalized to the excitation intensity at the QD site. This normalization is done by assuming an excitation Gaussian spot of 2.0 μm width centered at the CM and considering the intensity decrease at the QD sites. We end up with $\sqrt{I_{PL}(1)/I_{PL}(2)} = 0.58$. The CM electric field values at the QD sites normalized to the maximum value at the CM centre are calculated by three dimensional finite differences in the time domain (Lumerical code). The results are $E_1/E_{CM} = 3.2 \times 10^{-2}$ and $E_2/E_{CM} = 1.1 \times 10^{-2}$. We finally obtain from eq.(2) $g_1/g_2 \approx 10$. The main contribution to this large difference in g comes from the difference in the QD dipole orientation with respect to the CM polarization. In our case the detunings $\Delta_{1,2}$ are in the order of 1 meV, that together with $g_1 \approx 10g_2$, and taking a value of $g_1 \leq 0.05$ meV [11] lead to $\Delta_{1,2} \geq 20g_1$. Thus the effective coupling between the two QDs is $J \leq g_1/20$. This magnitude is much smaller than $|\Delta_1-\Delta_2|$ for typical values of the detunings. Therefore, population oscillations between the two qubits are expected to be slow enough to be easily detectable [17].

In conclusion, we have demonstrated the effective coupling between two distant QDs mediated by the electric field of a confined cavity mode. Individual coupling is observed by tuning each QD in resonance with the CM. The cavity mediated interaction is evidenced by the increased emission intensity of each QD upon resonant excitation in the excited state of the other one. The present results constitute an experimental step towards the realization of quantum logic operations using distant solid state qubits.

**Acknowledgements:**

The authors are indebted to E. Hu and R. Sabouni for helpful discussions. This work has been supported by research contracts of the Spanish Ministry of Education Grants MAT2008-01555/NAN, Consolider CSD 2006-19 and Naninpho-QD TEC2008-06756-C03-01, and the Community of Madrid Grant CAM S-0505-ESP-0200.

**Figure captions:**

**Figure 1**

(color online). (a,b) Photoluminescence spectra showing the exciton emission of QD1, QD2 and the CM for 7K (a) and 37K (b) for polarization parallel (V, black lines) and perpendicular (H, red line) to the CM. The inset in (b) shows a SEM image of the cavity structure with the approximate location of the QDs. (c) Energies of the QD1 and CM emission lines as a function of temperature. (d) Emission intensity of QD1 (black) and QD2 (red) as a function of detuning for V polarization. The intensities are normalized to the total emission (the sum of the QD and the CM intensities). (e,f) Polarization diagrams of the QD2 emission at 7K (e) and 37K (f). A polarization rotation is observed upon decreasing detuning.

**Figure 2**

(color online). Emission intensity dependence of QD1, QD2 and CM on excitation (vertical axis) and emission (horizontal axis) energies at 37K. The white arrow points the weak enhancement of QD2 PL emission upon p-excitation of QD1. The inset shows a diagram of the involved states and transitions. QD1,2-s,p indicate the first and second exciton states of each QD. CM is the energy of a cavity photon. Black solid arrows show the excitation (up) and emission (down) processes of QD1. The dashed arrow indicates the QD2 exciton transition induced by its coupling to QD1 through the cavity. Red arrows stand for the reciprocal case.

**Figure 3**

PLE spectra of QD1 and QD2 for V polarization at 7K (a) and 37K (b). Vertical lines are at p-state energies of QD1,2. The shoulders observed at the high energy side of the QD2 spectra and the asymmetric broadening towards lower energies of the QD1 ones indicate inter-dot coupling.

**Figure 4**

(color online). PLE spectra of QD2 for H polarization at different excitation intensities. The solid line is a Gaussian fit to the main peak (intra-dot excitation). The shoulder at higher energies is the contribution of the inter-dot cavity-mediated coupling to the emission (inter-dot excitation). Inset: dependence of the QD2 emission intensity on excitation power for intra-dot (red crosses) and inter-dot (blue triangles) excitation.

**Figure 1**

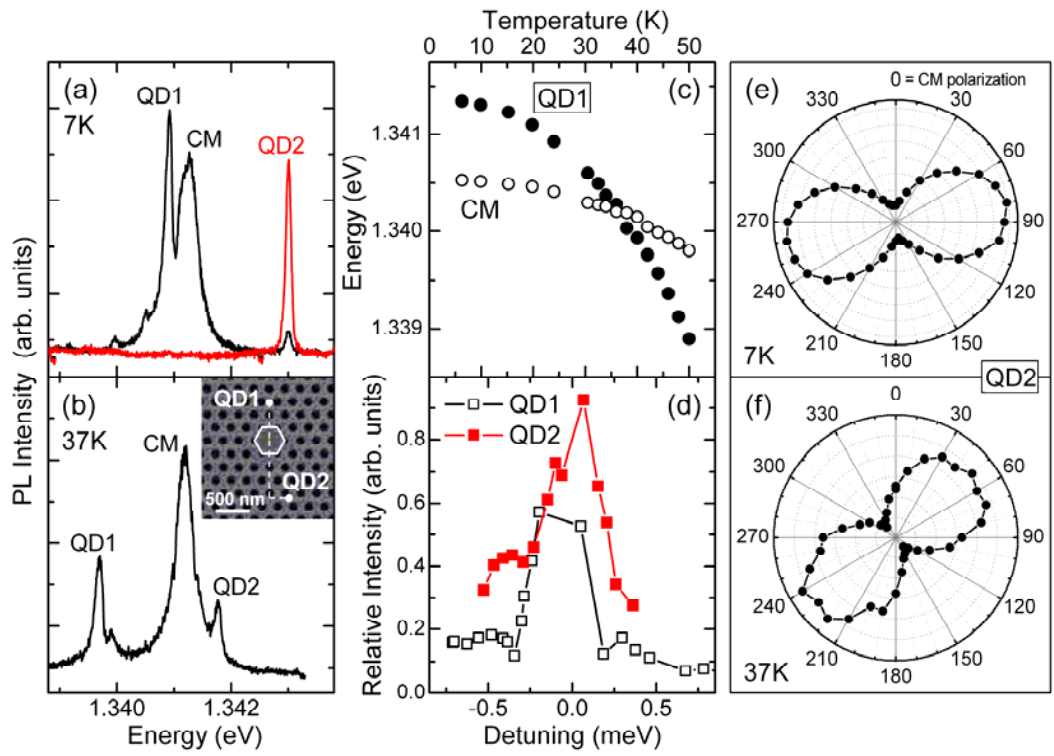

**Figure 2**

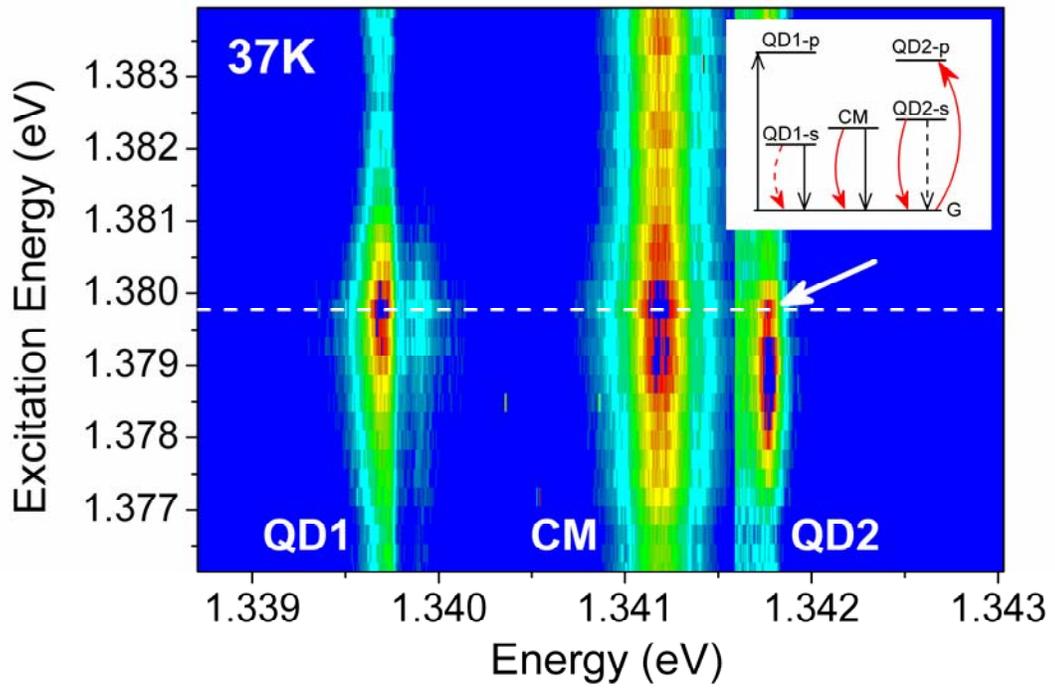

**Figure 3**

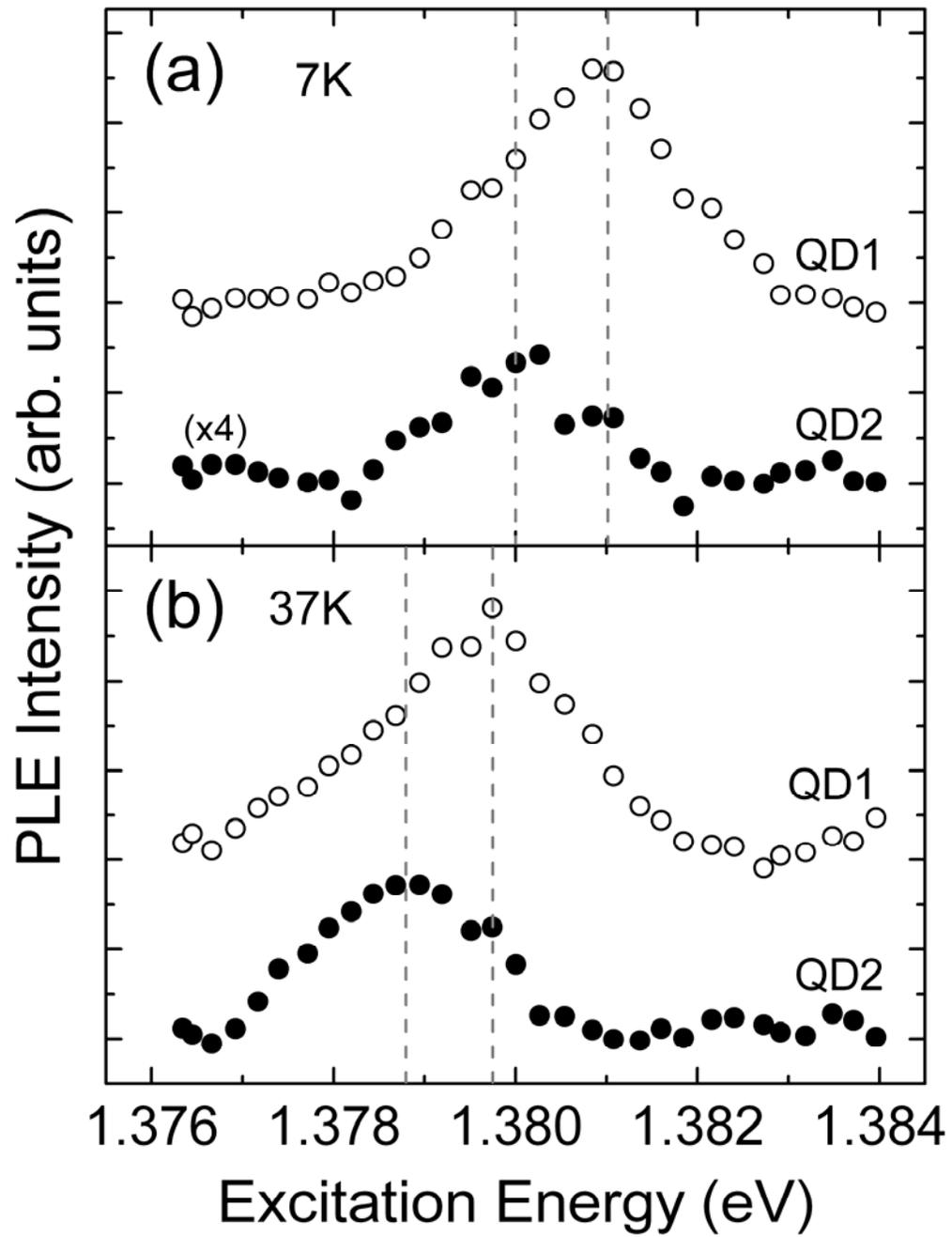

**Figure 4**

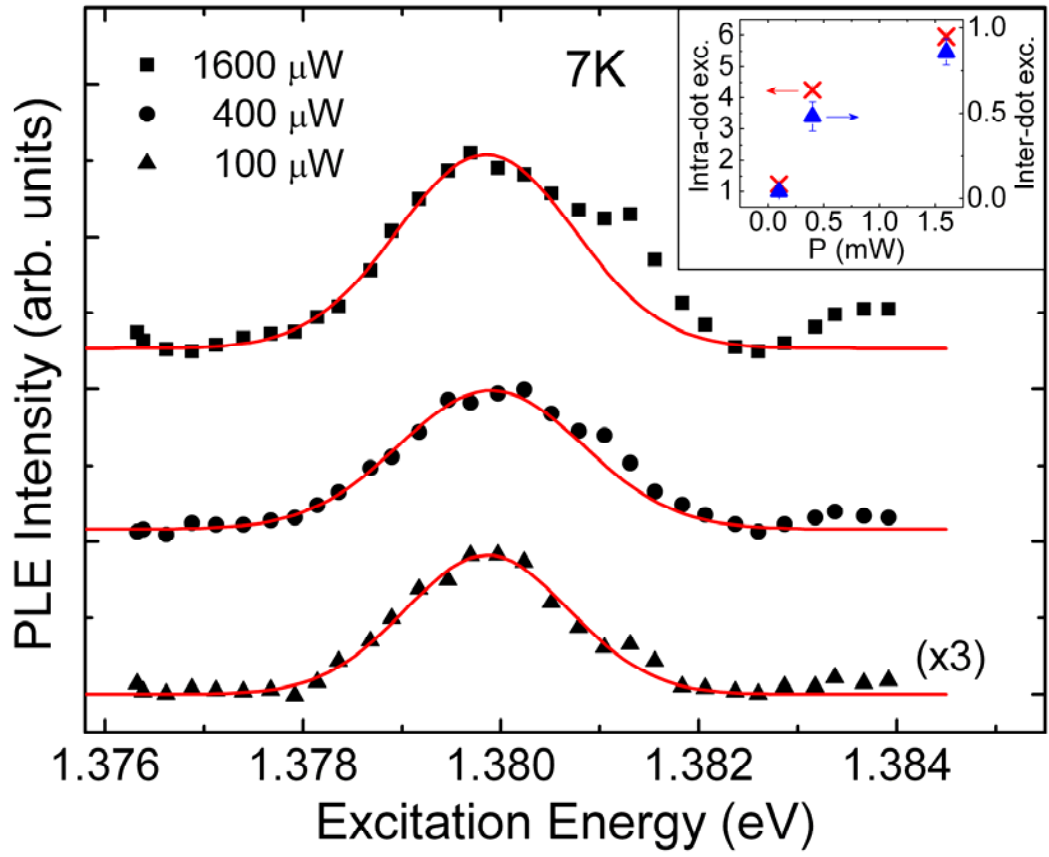